\input{aipcheck}

\documentclass[
    ,final            
  ]
  {aipproc}

\layoutstyle{6x9}

\newcommand{\sophia}{\textit{SOPHIA}}
\newcommand{\pks}{\textit{PKS~2155-304}}

\newcommand{\hess}{\textit{H.E.S.S.}}
\newcommand{\fermi}{\textit{Fermi}}
\usepackage[small]{caption}

\begin{document}

\author{M.Cerruti}{
  address={LUTH, Observatoire de Paris, CNRS, Université Paris Diderot; 5 Place
Jules Janssen, 92190 Meudon, France},
}

\author{A.Zech}{
  address={LUTH, Observatoire de Paris, CNRS, Université Paris Diderot; 5 Place
Jules Janssen, 92190 Meudon, France},
}
\author{C.Boisson}{
  address={LUTH, Observatoire de Paris, CNRS, Université Paris Diderot; 5 Place
Jules Janssen, 92190 Meudon, France},
}

\author{S.Inoue}{
  address={Institute for Cosmic Ray Research, University of Tokyo, Kashiwa, Japan}}
  
  \title{A mixed lepto-hadronic scenario for \pks}


\begin{abstract}
The models developed to describe the spectral energy distribution (SED) of blazars can be divided into leptonic or hadronic scenarios, according to the particles responsible for the high-energy component. We have developed a new stationary code which computes all the relevant leptonic and hadronic processes, permitting the study of both leptonic and hadronic scenarios in a consistent way. Interestingly, mixed lepto-hadronic scenarios (in which both components contribute to the high energy emission) naturally arise in this framework. We present the first application to the well known BL Lac object \pks. 
\end{abstract}

\maketitle


\section{Introduction}

In the framework of the unified active galactic nuclei (AGN) model, blazars are considered as radio-loud AGN whose relativistic jet is pointed close to the line-of-sight \citep{Urry95}. Their SED is characterised by two bumps, one peaking at low energies (infrared to X-rays), the other in $\gamma$-rays. While the first component is unanimously ascribed to synchrotron emission by a non-thermal population of electrons and positrons, the origin of the second component is still under discussion. In leptonic models it is ascribed to an inverse Compton process between the electrons and a low energy photon field (their own synchrotron emission, or external photons), while in hadronic models it originates from synchrotron emission by protons and secondary particles coming from $p$-$\gamma$ interactions.\\
We have developed a new stationary lepto-hadronic code which describes all the relevant process, permitting the study of both scenarios in a consistent framework, as well as interesting mixed lepto-hadronic scenarios. We first provide the details of the code, and we then present an application to the BL Lac object \pks.

\section{Description of the code}
   \label{sec2}
   
   \label{lep}
   The code we present is a direct evolution of the code developped by \citet{Kata01} 
   Several modifications and improvements have been applied to the original leptonic code: more details can be found in \citep{leha1}.
   The emitting region is assumed to be filled with a stationary population of relativistic electrons/positrons and protons. The proton energy distribution $N_p(\gamma_p)$ is described by a power-law function with index $\alpha_p$, defined above $\gamma_{p;min}$ and with an exponential cut-off at $\gamma_{p;Max}$. 
   
   
   The proton population in the emitting region interacts with the low energy photons through the photo-meson process 
   and the electron-positron pair production (Bethe-Heitler process).  
   The photo-meson interaction is evaluated using the publicly available Monte-Carlo code \sophia\ \citep{Sophia} which gives as output the distributions of the stable particles produced in the interaction. The low-energy photon field is represented by the synchrotron emission from the primary electron population in the jet, which represents by far the dominant component at low energies for high-frequency-peaked BL Lac objects as \pks. 
   The most important features of the computation of hadronic processes are the following:
   \begin{itemize}
   \item We first compute the synchrotron emission by protons.
   \item Before calling \sophia, the energy of the proton is modified in order to take into account the synchrotron losses following \citet{Mucke01}. 
   \item Photons coming from the $\pi^0$ decay and the synchrotron emission from $e^\pm$ coming from the $\pi^\pm$ channel can trigger an electro-magnetic cascade, supported by synchrotron emission. The stationary state of the cascade emission has been evaluated iteratively computing for each generation the stationary distribution of pairs and the associated synchrotron emission. 
   \item Within \sophia, we corrected for synchrotron losses the energy of kaons, pions and muons before decay.
   \item Within \sophia, we retrieved the muon ($\mu^\pm$) spectra \textit{before} their decay into electrons and positrons. The stationary spectrum is computed taking into account both decay and synchrotron cooling, and their synchrotron emission is then evaluated. 
   \end{itemize}

   We consider that the proton index is equal to the electron index before the break (under the assumption that the acceleration mechanism is similar), and we fix the maximum proton energy through a comparison between acceleration and cooling time-scale (see \citet{Mucke01}). With respect to a pure leptonic model, we add then only one free parameter in the code, the proton normalization factor.

\section{Modelling of \pks}
\label{sec3}

\begin{figure}
	   	\includegraphics[width=13.5cm, angle=0]{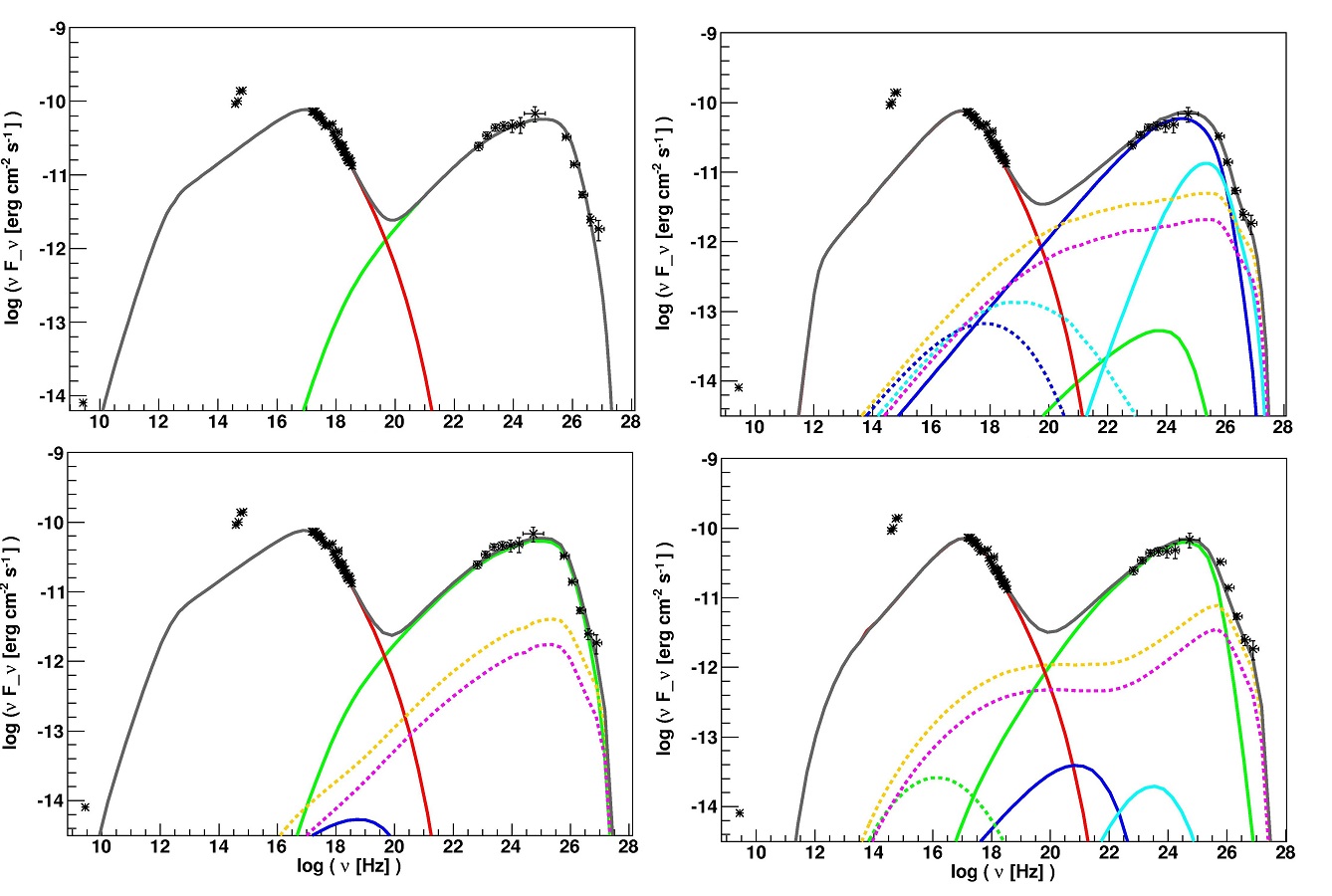}
	    \caption{\textit{Top left:} modelling of \pks\ in a synchrotron-self-Compton scenario. \textit{Top right:} modelling of \pks\ in a hadronic scenario. \textit{Bottom:} modelling of \pks\ in a lepto-hadronic scenario. The colour-code is as follow: red-primary electron synchrotron emission; green-inverse Compton emission; blue-proton synchrotron emission; light blue-muon synchrotron emission; yellow-photons from $\pi^0$ decay; pink-synchrotron emission from secondary $e^\pm$ coming from $\pi^\pm$ decay. The emission from the associated electro-magnetic cascade is shown by the dotted curves (same colour).}
	    \label{fig1}
  \end{figure}

As a first application we study the BL Lac object \pks, using in particular the data from the multiwavelength campaign of 2008 \citep{PKS}, in which the source was found to be in a low flux state.\\
We model the source in four different scenarios, shown in Fig. \ref{fig1} (the corresponding model parameters are given in Table \ref{lehatable1}): a standard synchrotron-self-Compton model, a hadronic model and two different lepto-hadronic models. The Doppler factor of the emitting region has been fixed at $30$ for all the models. For the SSC model we adopted a low magnetic field ($B=65$ mG), and a relatively large emitting region ($R\approx10^{16}$cm). In the hadronic model, in order to have a significant synchrotron emission from the protons, we need to increase the magnetic field up to $80$ G, reducing the emitting region by two order of magnitudes ($R\approx10^{14}$ cm) in order to fit the low energy bump with the primary electron synchrotron emission. Interestingly, the synchrotron emission from secondary particles ($\mu$ and cascades associated with $\gamma$ and $e^\pm$) coming from $p$-$\gamma$ interactions significantly contributes to the high-energy bump.\\
The first lepto-hadronic model uses the same parameters that were adopted for the SSC model, but we add protons in the emitting region. Given the low magnetic field, they do not contribute directly to the SED, but the emission from secondary pairs can play a role in the TeV energy range (in this case the power of the emitting region increases, reaching the Eddington luminosity of the source). The second lepto-hadronic model lies in between the SSC and the hadronic one, with a magnetic field of $5$ G. In this case the SSC component dominates only up to a few 10 GeV energies, while at higher energies the synchrotron emission by secondary pairs becomes the dominant process. In this case the \fermi\ and \hess\ data would then be ascribed to two different processes.\\
An interesting result is that the presence of hadrons affects the $\gamma$-ray emission, inducing a hardening of the TeV spectrum (that can be seen in both the hadronic model and the first lepto-hadronic model). This effect can be tested by the next-generation Cherenkov telescope \textit{CTA}. Another important feature is that the cascade component tends to fill the gap between the two bumps much more than in pure SSC models. The upcoming hard-X-ray satellites \textit{NuSTAR} and \textit{Astro-H} will put new constraints on this part of the spectrum.

   \begin{table}[ht!]
   \begin{tabular}{l c c c c}
		\hline
		\hline
		& SSC & Hadronic & Mixed 1 & Mixed 2\\
		\hline
		$\theta$ & $1^\circ$ & $1^\circ$ & $1^\circ$ & $1^\circ$\\
		$\delta$ & $30$ & $30$ & $30$ & $30$\\
		$\Gamma_{bulk}$ & $16$ & $16$ & $16$ & $16$\\
		\hline
		$\gamma_{e,min}$& $5\times10^2$ & $1$ & $5\times10^2$ & $1\times10^2$\\
		$\gamma_{e,break}$& $1.5\times10^5$ & $4\times10^3$ & $1.5\times10^5$&$1.7\times10^4$\\
		$\gamma_{e,max}$& $3\times10^6$& $6\times10^4$& $2.5\times10^6$ &$3\times10^5$\\
		$\alpha_{e,1}$ &$2.4$&$2.0$&$2.4$ &$2.0$\\
		$\alpha_{e,2}$ &$4.32$&$4.32$&$4.32$ &$4.32$\\
		$K_e\ \textrm{[cm}^{\textrm{-3}}\textrm{]}$ & $7.5\times10^4$ & $6.0\times10^2$&$6.3\times10^4$ &$6\times10^5$\\
		$u_e\ \textrm{[erg cm}^{\textrm{-3}}\textrm{]}$ & $5.9\times10^{-3}$ & $2.2\times10^{-3}$& $5.1\times10^{-3}$&$1.4$\\
		\hline
		$\gamma_{p,min}$& -& $1$& $10^5$ &$1$\\
		$\gamma_{p,max}$& -& $1\times10^9$& $8\times10^7$ &$6\times10^7$\\
		$\alpha_{p}$ & -& $2.0$& $2.4$&$2.0$\\
		$\eta = K_p/K_e$ & -& $20$& $10$ &$1.0$\\
		$u_p\ \textrm{[erg cm}^{\textrm{-3}}\textrm{]}$ & -& $3.7\times10^2$& $22$ &$1.6\times10^4$\\
		\hline
		$R_{src}$ [cm] & $1.7\times10^{16}$& $5.2\times10^{14}$&$1.8\times10^{16}$ &$2\times10^{14}$\\
		\hline
		$B$ [G] & $0.065$& $80$& $0.065$&$5$\\
		$u_B\ \textrm{[erg cm}^{\textrm{-3}}\textrm{]}$ & $1.7\times10^{-4}$& $2.5\times10^2$& $1.7\times10^{-4}$&$1.0$\\
		\hline
		$u_e/u_B$ & $35$ & $8.5\times10^{-6}$ & $29$&$1.4$\\
   	$u_p/u_B$ &  - & $1.5$ & $1.3\times10^{5}$ &$1.6\times10^{4}$\\
		$L_{jet}\ \textrm{[erg s}^{\textrm{-1}}\textrm{]}$& $4.3\times10^{43}$& $4.2\times10^{45}$&$1.7\times10^{47}$ &$1.5\times10^{46}$\\
		\hline
		\hline			
		\end{tabular}
		\caption{Parameters used for the SSC and the hadronic modelling of \pks. The luminosity of the source has been calculated as $L_{jet}=\pi R_{src}^2c\Gamma_{bulk}^2(u_B+u_e+u_p)$. For the SSC scenario, we included the term $u_{p,cold}$, which represents the contribution of cold protons in the emitting region, estimated following \citet{Sikora09}.}
		\label{lehatable1}
		\end{table}

 \bibliographystyle{aipproc}
 \bibliography{leha_paper}
\end{document}